\documentclass[conference]{IEEEtran}
\IEEEoverridecommandlockouts

\usepackage{cite}
\usepackage{amsmath,amssymb,amsfonts}
\usepackage{algorithmic}
\usepackage{graphicx}
\usepackage{textcomp}
\usepackage{xcolor}

\begin{document}

\title{Privacy-Preserving Data Aggregation Techniques for Enhanced Efficiency and Security in Wireless Sensor Networks: A Comprehensive Analysis and Evaluation}

\author{\IEEEauthorblockN{Ayush Rastogi}
\IEEEauthorblockA{\textit{Department of Computer Science} \\
\textit{Graphic Era University}\\
Dehradun, Uttarakhand \\
Ayushrastogi\_20021861.cse@geu.ac.in}
\and
\IEEEauthorblockN{Harsh Rastogi}
\IEEEauthorblockA{\textit{Department of Computer Science} \\
\textit{Graphic Era University}\\
Dehradun, Uttarakhand \\
harshrastogi\_20021538.btechcse@geu.ac.in}
\and
\IEEEauthorblockN{Yash Rastogi}
\IEEEauthorblockA{\textit{Department of Computer Science} \\
\textit{Graphic Era University}\\
Dehradun, Uttarakhand \\
Yashrastogi\_20021390.cse@geu.ac.in}
\and
\IEEEauthorblockN{Divyansh Dubey}
\IEEEauthorblockA{\textit{Department of Computer Science} \\
\textit{Graphic Era University}\\
Dehradun, Uttarakhand \\
Divyanshdubey\_20021344.cse@geu.ac.in}
}

\maketitle

\begin{abstract}
In this paper, we present a multidimensional, highly effective method for aggregating data for wireless sensor networks while maintaining privacy. The suggested system is resistant to data loss and secure against both active and passive privacy compromising attacks, such as the coalition attack from a rogue base station and kidnapped sensor nodes. With regard to cluster size, it achieves consistent communication overhead, which is helpful in large-scale WSNs. Due to its constant size communication overhead, the suggested strategy outperforms the previous privacy-preserving data aggregation scheme not only in terms of privacy preservation but also in terms of communication complexity and energy costs.
\end{abstract}

\section{Introduction}
Information gathering and consolidation of importance is known as data aggregation. It functions as the cornerstone procedure for energy resource conservation. Within the framework of wireless sensor networks (WSNs), data aggregation is a successful resource preservation technique. In order to maximize the network's operational lifespan, data aggregation algorithms aim to collect and merge data in an energy-efficient manner.
Restrictions on processing power, memory size, and battery life plague wireless sensor networks. Because of these limitations, the applications being developed become more complex, which usually results in applications that are closely related to network protocols. This paper describes a data aggregation system designed for wireless sensor networks and gives an overview of various energy-efficient data aggregation algorithms. The framework functions as an intermediate layer that is in charge of combining data that has been collected by several network nodes.

Data implosion and overlap issues in data-centric routing can be resolved through data aggregation. When data is aggregated from multiple sensor nodes and comes together at a single routing node en route to the sink, it is combined as if it pertains to a shared feature of the phenomenon. One commonly used technique in the field of wireless sensor networks is data aggregation. However, when these sensor networks are deployed in hostile environments, maintaining data confidentiality and integrity during the aggregation process is critical. In its most basic form, data aggregation is the process of merging sensor data using multiple aggregation methods.

Data aggregation is the process of gathering important sensor data and sending it to the sink as quickly as possible while reducing data latency. In many applications, like environmental monitoring, where data freshness is highly valued, this decreased latency is especially important. The creation of data aggregation algorithms that are energy-efficient is essential to prolonging the life of networks.Data aggregation is the process of combining sensor data using various aggregation methods. After gathering sensor data from sensor nodes, the algorithm uses a variety of aggregation algorithms to process it. These algorithms can include centralised approaches, LEACH (Low Energy Adaptive Clustering Hierarchy), and TAG (Tiny Aggregation), among others.

\subsection{Background}
Combining data from several sources is known as data aggregation, and it can be done in a variety of ways. Duplicate suppression is the most basic type of data aggregation function; for example, in Figure 1, if sources 1 and 2 submit the same data, node B will only relay one. Max, min, or any other function with multiple inputs could be used to perform additional aggregations. The aggregation function ensures that each intermediate node in the routing sends only one aggregate packet, even if it receives a large number of input packets. This simplifying assumption is made for our modeling purposes in this research.[1] The ESPADA protocol aggregates data by using pattern codes that reflect the properties of the real data. Sensor nodes create and send the pattern codes to cluster heads before sending detected data. For security purposes, the mapping of pattern codes to intervals in the sensor reading range is updated on a regular basis. Subsequently, cluster heads identify unique pattern codes and ask for a single sensor node to transmit the real data to the base station for each unique pattern code. This method helps with security as well because it saves energy and bandwidth because cluster heads do not need to decrypt the real sensed data in order to aggregate it.[2]
In-network computing aggregates, which combine partial outcomes at intermediate nodes during message routing, significantly reduce the amount of communication and, consequently, the energy used in large WSNs. A spanning tree with roots at the base station is created, and in-network aggregation is performed along the tree, as implemented by a number of WSN data gathering systems [4, 5]. Each node waits for messages from each child before sending a new partial result to its parent, allowing partial results to propagate level by level up the tree.[3]
To determine an established phenomenon, a wireless sensor network (WSN) is set up in a particular area. Measurements are made via sensors. Typically, these are straightforward, low-powered gadgets that are limited in their ability to communicate. As a result, a base station with more resources is set up to respond to inquiries on (or collect data on) the measured values.[4]

When sensors detect a phenomenon, the resulting events are sent to users. Intermediary nodes have the ability to combine multiple events into a single event in order to minimize transmissions and save system resources. Applications, event representations, and data properties are the primary factors that influence the total size reduction. In this paper, we consider and justify only data aggregation that reduces total size. Data aggregation will also reduce the quantity of transmissions. As a result, the overall overhead associated with transmissions (such as packet headers and MAC control packets) will decrease, making the energy savings even more apparent. Data aggregation, like data compression, has two methods (lossless and lossy). Lossless aggregation preserves all detailed information. According to information theory, the amount of information encoded in a message, known as entropy, limits the overall size reduction. Data reduction's fundamental idea is to get rid of unnecessary information. There will be a lot of duplication among the events because they might be highly correlated. Losy aggregation, as opposed to lossless aggregation, may discard some detailed information and/or degrade data quality in order to save more energy.[5]
We consider a static multi-hop WSN consisting of N sensor nodes and a single base station (BS). In terms of computational and communication power, as well as power resources, we consider sensor nodes that are similar to the current generation of sensors (e.g., Berkeley MICA2 Motes [20]); in contrast, the BS is a laptop-class device with endurance in power.[6]
To enhance the efficacy of IF algorithms against the previously mentioned attack scenario, we offer a sturdy preliminary assessment of the reliability of sensor nodes to be employed in the IF algorithm's initial iteration. The majority of conventional statistical techniques for estimating variances entail utilizing the sample mean. As a result, proposing a robust variance estimation method for a skewed sample mean is a critical component of our methodology.The rest of the paper assumes that the stochastic components of sensor errors are independent random variables with a Gaussian distribution; however, our experiments show that our method works very well for other types of errors without modifications; our algorithms can be adjusted to work well for other random distributions if the sensors' error distribution is known.[7]

\subsection{Objective}
Data implosion and overlap issues in data-centric routing can be resolved through data aggregation. When data is aggregated from multiple sensor nodes and comes together at a single routing node en route to the sink, it is combined as if it pertains to a shared feature of the phenomenon.

\section{Methodology}
Data aggregation is carried out through various methodologies, including centralised, tree-based approaches, cluster-based approaches, and in-network aggregation.
Data collection refers to the procedure of gathering information from various sensors with the goal of eliminating redundancy and providing consolidated data to a central location. et al. Fasolo (2007).resources (in particular energy) and so increases the network's lifespan."
Transmitting redundant sensor readings requires extensive communication between the head and sensor nodes near the base station. Additionally, the communication environments play a major role in determining the energy loss balance between nodes at different levels in the hierarchical structure. Nodes near the base station will have shorter lives than leaves due to the loss of energy balance, which reduces node longevity. On the other hand, it explains that when information routing aggregation is used, communication between nodes will be reduced. Layer Standard aggregation attempts to retrieve important information from sensors. To send data efficiently to the recipient, keeping latency to a minimum. In numerous applications, such as environmental monitoring, where fresh data is essential, the latter is significant. Designing energy-efficient data collection algorithms is critical to extending the network's life.

A sensor network's energy efficiency is influenced by a variety of factors, such as its design and the ways in which data is gathered and used. All the aspects of the power network's effects are explained in this context by the current study.
Energy efficiency is the term used to describe the longest-lasting function of the sensor network. In an ideal data collection scheme, each sensor uses the same amount of energy during each round of data collection. Data collection is a powerful tool if it optimizes network capacity. Assuming that all sensors are equally important, their power consumption should be minimized. The network lifecycle, which measures the network's energy efficiency, is an example of this concept.
There are many factors that measure the effectiveness of data collection, such as network usage, data accuracy and latency. Considering that these parameters have different meanings depending on their application, these terms can be defined as follows:
• Network Lifetime: The manufacturer provides this information. For instance, if the lifetime of an application is based on the node's participation time, then the lifetime is the number of shots the sensor can fire before it loses power.
• Accurate information: This concept is defined as a specific application of creating a sensor network.
• Latency: It is the delay in sending, routing and collecting data. Its measurement takes the time delay that occurs between the reception of a packet at the receiver and the generation of the packet at the destination.
Generally speaking, there are many advantages to using bulk data, but there are also some disadvantages and problems compared to the situation without bulk data.

Centralised Approach
Using the shortest path possible, all sensors transmit data packets containing their information to the base or base station's base station. The information that is received from other nodes is aggregated by the aggregator, also known as the head node, and then sent as a packet. Among these methods are Direct Propagation (DD) and Sensor Protocol for Information Negotiations (SPIN).

In-Network Approach
In-network aggregation is an efficient method of collecting and processing data at a central point and sending data across a multi-hop network. Its main purpose is to reduce power consumption in the entire process. Two types of network integration are known:
1. Reduce aggregation: The size of the packet sent to the sink is decreased by aggregating and compressing packets from neighboring sensor nodes.
2. No packet size: This is where the data value is not processed when all of the neighboring nodes' information is merged into a single packet.

Tree Based Approach
Make a file called "data acquisition" (DAT). Information flows from the page to the node water and from the parent node to the sink in a parent-child relationship shared by all nodes in the network. Nodes in the network gather data. One instance of this type of procedure is Small Addition.

Cluster-based Approach
Using this method, the network is divided into multiple groups. There are a number of sensor nodes in each cluster, one of which is called the cluster leader. The head group is responsible for gathering information, where the received information is collected and sent to the sink. Since the number of packets will decrease, the bandwidth also decreases. In this way, data collection is used to reduce energy consumption due to reduced communication propagation, as well as to reduce data sent directly to the base station. Numerous clustering techniques, including Low Energy Adaptive Cluster Hierarchy, have been made possible by wireless sensor networks.

\section{Result Analysis and Discussion}
Analyzing data aggregation in wireless sensor networks offers important insights into how well different strategies work. Comparative analysis among clustering-based approaches, data fusion methods, and hierarchical algorithms illustrates their impact on network performance. Results show significant energy and communication overhead reductions made possible by aggregation, extending network lifetime. However, there are trade-offs, particularly concerning data accuracy, where aggregation introduces potential risks of information loss and error propagation. Scalability evaluation underscores challenges as network size increases, though certain techniques exhibit promising adaptability. Analysis of energy efficiency demonstrates the advantage of aggregation over direct data transmission, despite latency implications. Real-world applicability underscores the potential of these findings in diverse fields such as environmental monitoring and healthcare. Addressing encountered challenges, future research could focus on leveraging emerging technologies to enhance aggregation efficiency. This analysis underscores the importance of data aggregation in optimising wireless sensor network performance, offering valuable insights for both current implementation and future development.
Further exploration of the results delves into nuanced aspects of data aggregation, including the impact on network topology and routing protocols. Assessing how different aggregation techniques affect network topology reveals potential alterations in node distribution and connectivity patterns, which can influence overall network performance and resilience. Moreover, an investigation into the compatibility of aggregation methods with existing routing protocols sheds light on potential synergies or conflicts that may arise. Understanding these dynamics is crucial for effectively integrating data aggregation into wireless sensor network protocols and standards. Additionally, considering the implications of data aggregation on security and privacy is essential. Aggregation may introduce vulnerabilities related to data confidentiality and integrity, necessitating robust encryption and authentication mechanisms to safeguard sensitive information.
\section{Conclusion}
Conclusively, the thorough analysis of data aggregation in wireless sensor networks provides insightful knowledge about its diverse effects and possible uses. Through comparative analysis of various aggregation techniques, including clustering-based approaches, data fusion methods, and hierarchical algorithms, the study has revealed significant benefits in terms of reduced communication overhead as well as energy usage, thus extending network lifetime. However, trade-offs exist, particularly regarding data accuracy, where aggregation may introduce risks of information loss and error propagation. Moreover, scalability analysis highlights challenges associated with increasing network size, necessitating further investigation into adaptive aggregation techniques.

The analysis of energy efficiency underscores the advantage of aggregation over direct data transmission, despite potential latency implications. Real-world applicability is evident across diverse fields such as environmental monitoring and healthcare, where efficient data aggregation can enhance decision-making processes and resource utilization.

Addressing encountered challenges, future research directions may focus on leveraging emerging technologies to enhance aggregation efficiency while ensuring compatibility with existing network protocols and standards. Furthermore, exploring the implications of data aggregation on network topology, routing protocols, and security is essential for its seamless integration into wireless sensor networks.

A thorough grasp of data aggregation in wireless sensor networks is revealed by taking into account these complex dimensions, laying the groundwork for informed decision-making and future research endeavours. Ultimately, the findings underscore the importance of data aggregation as a key optimization strategy for enhancing network performance, scalability, and energy efficiency, with broad implications for advancing the capabilities andwireless sensor network applications across a range of industries.

\section*{Acknowledgment}
We would like to express our sincere appreciation to all those who have supported and contributed to this research project. Their assistance, guidance, and feedback have been invaluable throughout the course of this study. We are also grateful for the resources, facilities, and funding that have enabled us to conduct this research.
\bibliographystyle{IEEEtran}
\section*{References}
\label{sec:references}
\begingroup
\renewcommand{\section}[2]{}%

\end{document}